# "GENERAL THEORY OF A PARTICLE MECHANICS" ARISING FROM A FRACTAL SURFACE

Alexander P. Yefremov[1]

Institute of Gravitation and Cosmology of Peoples' Friendship University of Russia

Abstract. The logical line is traced of formulation of theory of mechanics founded on the basic correlations of mathematics of hypercomplex numbers and associated geometric images. Namely, it is shown that the physical equations of quantum, classical and relativistic mechanics can be regarded as mathematical consequences of a single condition of stability of exceptional algebras of real, complex and quaternion numbers under transformations of primitive constituents of their units and elements. In the course of the study a notion of basic fractal surface underlying the physical three-dimensional space is introduces, and an original geometric treatment (admitting visualization) of some formerly considered abstract functions (mechanical action, space-time interval) are suggested.

## 1. Introduction

In the previous paper [1] a logical line describing growth of different branches of contemporary mechanics from mathematical roots was verbally exposed. In the present paper this logic is traced using purely math language finely demonstrating surprising results. The method used here implies a fine analysis of primitive elements constituting the basic structures of exceptional associative algebras (and other poly-number algebras). The simplest transformations of the elements induce involvement of a series of conditions (equations) providing the algebras stability; these pure math equations are found equivalent to laws of different branches of mechanics, and become them exactly when rewritten in physical units (instead of abstract magnitudes). This unique logical line leads consequently to equations of quantum, classical and relativistic mechanics traditionally thought of as belonging to somewhat separate theories. Two types of main object of the theory, of a particle, indispensably emerge, a fractal one, made up from primitive elements, and a geometric one observed in the physical space. By the way some traditionally abstract functions (mechanical action, space-time interval) acquire an original geometric sense.

The paper is organized as following. Section 2 contains a short review of the involved algebras, their units represented by matrices. Section 3 outlines the principal stages and main points of the "general theory of mechanics". In particular, transformations of a fractal surface spoiling the algebras' basis are analyzed, and a condition of the algebras stability is introduces then decaying into a series of fractal and geometric equations of mechanics (written in purely math and physical units). Respective particle's model gives birth to a version of special relativity (with a helix-type Minkowski diagram) and to a general relativity-type geodesic equation. Compact discussion in section 4 concludes the study.

## 2. Algebras involved

We start with *bi-quaternions*, hypercomplex numbers of the form [2]

$$b \equiv x + iy + (u_1 + iw_1)\mathbf{q}_1 + (u_2 + iw_2)\mathbf{q}_2 + (u_3 + iw_3)\mathbf{q}_3 \equiv x + iy + (u_k + iw_k)\mathbf{q}_k; \tag{1}$$

here $x, y, u_k, w_k$ are real numbers, the scalar part $x + iy$ has real unit factor 1 (traditionally not shown), $\mathbf{q}_k$ are three vector units, all four units satisfying the multiplication law

---
[1] Address: Russia, 117198 Moscow, Miklukho-Maklaya str., 6; e-mail: a.yefremov@rudn.ru

$$\mathbf{q}_k 1 = 1 \mathbf{q}_k, \quad \mathbf{q}_k \mathbf{q}_n = -\delta_{kn} + \varepsilon_{knm} \mathbf{q}_m, \tag{2}$$

$\delta_{kn}$, $\varepsilon_{knm}$ are 3D Kroneker and Levi-Civita symbols, summation in repeated (3D small Latin) indices is assumed. The numbers of the type (1) constitute the largest algebra in question here; according to the law (2) the algebra of bi-quaternions is non-commutative, but it is associative in multiplication. It is well known that the norm of a bi-quaternion in general is not defined as a positive real number, so this algebra has division defects, in particular zero dividers.

If in Eq.(1) $x \neq 0, u_k \neq 0$ while all imaginary components vanish $y = w_k = 0$, then one arrives to a q*uaternion*, hypercomplex number of the form $q \equiv x + u_k \mathbf{q}_k$, the units $(1, \mathbf{q}_k)$ obeying the law (2); so the quaternions can be regarded as a (4-unit) section of the set of bi-quaternions. The norm and the modulus of a quaternion are well defined $\|q\|^2 \equiv q\bar{q} = (x + u_k \mathbf{q}_k)(x - u_k \mathbf{q}_k) = x^2 + u_k u_k$, $|q| \equiv (q\bar{q})^{1/2} = \sqrt{x^2 + u_k u_k}$, hence the inverse (left and right) number and division exist. The Frobenius theorem proves that the algebra of quaternions is the last in dimension associative, though non-commutative, division algebra (non-commutative ring).

If in Eq.(1) e.g. only $x \neq 0, w/2 \equiv w_1 = u_2 \neq 0$, while all other components vanish, then one obtains a "more narrow" (2-unit) section of bi-quaternions, the set of "exotic" *dual numbers* [3] of the form $d \equiv x + w(i\mathbf{q}_1 + \mathbf{q}_2)/2 \equiv x + w\boldsymbol{\varepsilon}$, its vector unit having zero norm $\boldsymbol{\varepsilon}^2 = 0$. The algebra of dual numbers is associative and commutative (according to Eq.(2) $1\boldsymbol{\varepsilon} = \boldsymbol{\varepsilon} 1$), but it includes zero dividers since the norm of a dual number depends only on the scalar part $\|d\|^2 \equiv d\bar{d} = (x + w\boldsymbol{\varepsilon})(x - w\boldsymbol{\varepsilon}) = x^2$, and a pure "imaginary" number has zero norm. It is evident that a dual number can be obtained from Eq.(1) in various ways, e.g. if only $x \neq 0, w/2 \equiv w_2 = u_3 \neq 0$, all other components vanishing or $x \neq 0, w/2 \equiv w_3 = u_1 \neq 0$, ect.

If in Eq.(1) e.g. only $x \neq 0, w \equiv w_1 \neq 0$, while the rest of the components vanish, then one meets another 2-unit section of bi-quaternions, set of *double numbers* (*split complex numbers*) [4] of the form $h \equiv x + wi\mathbf{q}_1 \equiv x + w\mathbf{p}$, its units commuting, and square of vector unit being equal to the real unit $\mathbf{p}^2 = 1$. The norm of $h$ is not well defined $\|h\|^2 \equiv (x + w\mathbf{p})(x - w\mathbf{p}) = x^2 - w^2$, so the commutative and associative algebra of dual numbers also has zero dividers. It is evident that a double number can be obtained from Eq.(1) in various ways.

If in Eq.(1) e.g. only $x \neq 0, u \equiv u_2 \neq 0$, while all other components vanish, then the last specific 2-unit section of bi-quaternions emerges, that of *complex numbers* $z \equiv x + u\mathbf{q}_2 \equiv x + u\mathbf{i}$. It is evident that various ways to select a complex number from Eq.(1) exist, in particular $z \equiv x + iy$ (with the traditional scalar imaginary unit), all these representations algebraically equivalent.

The simplest section of the bi-quaternion set is the 1-unit set of *real numbers* $x$, all other components in Eq.(1) being zero.

Algebras of real, complex and quaternion numbers are referred to as exceptional ones since only their elements (and 8-unit *octonion* numbers) satisfy the "square identities", the norm definition of a two elements product; e.g. for two quaternions $q_1$, $q_2$ it is $\|q_1 q_2\|^2 = \|q_1\|^2 \|q_2\|^2$. But the algebra of octonions is not associative, the property alien to known physical magnitudes; so octonions will be not considered here.

If the (bi-) quaternion algebra units are "canonically" represented by the $2 \times 2$ - matrices

$$1 = \begin{pmatrix} 1 & 0 \\ 0 & 1 \end{pmatrix}, \quad \mathbf{q}_{\tilde{1}} = -i\begin{pmatrix} 0 & 1 \\ 1 & 0 \end{pmatrix}, \quad \mathbf{q}_{\tilde{2}} = -i\begin{pmatrix} 0 & -i \\ i & 0 \end{pmatrix}, \quad \mathbf{q}_{\tilde{3}} = -i\begin{pmatrix} 1 & 0 \\ 0 & -1 \end{pmatrix}, \tag{3}$$

then the multiplication law (2) is satisfied identically; the other units then can be e.g. chosen as

$$\boldsymbol{\varepsilon} = (i\mathbf{q}_{\tilde{1}} + \mathbf{q}_{\tilde{2}})/2 = \begin{pmatrix} 0 & 0 \\ 1 & 0 \end{pmatrix}, \quad \mathbf{p} \equiv i\mathbf{q}_{\tilde{1}} = \begin{pmatrix} 0 & 1 \\ 1 & 0 \end{pmatrix}, \quad \mathbf{i} = \mathbf{q}_{\tilde{2}} = \begin{pmatrix} 0 & -1 \\ 1 & 0 \end{pmatrix}.$$

But many other representations exist since the basic law (2) evidently holds for the transformed units

$$\mathbf{q}_k = S\mathbf{q}_{\tilde{k}}S^{-1}, \tag{4}$$

matrices of the transformation forming the spinor group[2] $S \in SL(2,C)$ [5]. We stress that after any such transformation the scalar unit remains a unit matrix, while the vector units may become a multi-component function of many parameters. Note also that representations of the units $(1, \mathbf{q}_k)$ by matrices of the rank $2^N$ ($N$ is a natural number) can be readily introduced.

## 3. General theory of mechanics

### 3.1. Structures of algebraic units

(i) *All basic units of associative algebras (of real, complex, quaternion, double, dual and bi-quaternion numbers) can be regarded as matrices composed of a single dyad (a couple of 2D unit orthogonal vectors) on a fractal[3] surface.*

Let two 2D vectors $\psi^+$, $\psi^-$ (two-component column matrices) compose a local orthonormal basis (dyad) on a surface with a symmetric metric $g$, i.e. in a surface point $g\psi^+\psi^+ = 1$, $g\psi^-\psi^- = 1$, $g\psi^+\psi^- = g\psi^-\psi^+ = 0$; or with introduction of the covectors $\varphi^+ \equiv g\psi^+$, $\varphi^- \equiv g\psi^-$ (two-component row matrices) the dyad's definition is

$$\varphi^{\pm}\psi^{\pm} = 1, \quad \varphi^{\pm}\psi^{\mp} = 0. \tag{5}$$

Due to Eq.(5) the following linear combinations of direct products (tensor squares) of the set $\psi^{\pm}, \varphi^{\pm}$ satisfy the law (2) for the units $(1, \mathbf{q}_k)$ [6]

$$\psi^+\varphi^+ + \psi^-\varphi^- = 1, \quad -i(\psi^+\varphi^- + \psi^-\varphi^+) = \mathbf{q}_1, \quad \psi^+\varphi^- - \psi^-\varphi^+ = \mathbf{q}_2, \quad i(\psi^+\varphi^+ - \psi^-\varphi^-) = \mathbf{q}_3; \tag{6}$$

the other vector units can be chosen as (other combinations are possible)

$$\psi^-\varphi^+ = \boldsymbol{\varepsilon}, \quad \psi^+\varphi^- + \psi^-\varphi^+ = \mathbf{p}, \quad \psi^+\varphi^- - \psi^-\varphi^+ = \mathbf{i}.$$

(ii) *The metric of the dyad's local vicinity (2D-cell) behaves as the real unit of all involved algebras, while the three vector units $\mathbf{q}_k$ behave as a Cartesian frame in a 3D space.*

A domain of the fractal surface in vicinity of the dyad's origin together with part of tangent plane having the metric $\delta_{MN} = \delta^{MN} = \delta_M^N$ (unit $2\times 2$-matrix) will be called a "2D-cell". So we can identify the scalar unit $1$ with the metric of a 2D-cell, the metric of the fractal surface structured by the dyad covectors $g = \varphi^+\varphi^+ + \varphi^-\varphi^-$. The quaternion vector units $\mathbf{q}_k$ from the times of Hamilton [7] are known to be geometrically identified with a frame initiating a Cartesian system of reference in a 3D space often associated with the physical space.

---

[2] The law (2) is as well invariant under *SO*(3,C) transformations of the units, this group of rotations is twice covered by the reflection group *SL*(2,C).
[3] "Fractal" means here that a line dimension on the surface is ½ of that of 3D geometric (physical) space.

(iii) *Dimension of a line on 2D-cell's (e.g. a dyad vector length) is a square roots from dimension of a line in 3D space (e.g. a vector unit length); from the 3D space's viewpoint the dyad vectors are spinors.*

Eqs.(6) demonstrate that the dyad vectors (covectors) may be regarded as specific square roots of the units $(1, \mathbf{q}_k)$, but the single dyad is sufficient to build all units. Also note that the vectors (covectors) of this dyad are right and left eigenfunctions of the unit $\mathbf{q}_3$ with eigenvalues $\pm i$

$$\mathbf{q}_3 \psi^\pm = \pm i \psi^\pm , \quad \varphi^\pm \mathbf{q}_3 = \pm i \varphi^\pm , \tag{7}$$

(that is why the parity indicators $\pm$ arise). Eigenfunctions of the simplest operator $\mathbf{q}_{\tilde{3}}$ from Eqs.(3a) are

$$\tilde{\psi}^+ = \begin{pmatrix} 0 \\ 1 \end{pmatrix}, \quad \tilde{\psi}^- = \begin{pmatrix} 1 \\ 0 \end{pmatrix}, \quad \tilde{\varphi}^+ = (0 \quad 1), \quad \tilde{\varphi}^- = (1 \quad 0). \tag{8}$$

We emphasize that the transformations of the dyad constituents by matrices $S \in SL(2,C)$

$$\psi^\pm = S \tilde{\psi}^\pm , \quad \varphi^\pm = \tilde{\varphi}^\pm S^{-1} \tag{9}$$

should be considered the prior ones, since they induce the transformations (4) of 3D units leaving the multiplication law (20) intact; so the dyad vectors (covectors) $\psi^\pm$, $\varphi^\pm$ are spinors.

(iv) *If 3D space and objects in it are traditionally attributed to "geometry", then the fractal surface and objects on it (2D-cell, dyad vectors) may be related to "pregeometry".*

We dare remind that the notion of pregeometry was introduced by Wheeler in attempt to find a plausible image of a space where functions and operators of quantum mechanics act [8].

3.2. Transformations of a 2D-cell

(v) *The 2D-cell's area can be "pumped over" from real to imaginary sector with a certain phase, this "flickering" does not change the metric, but respective 3D frame rotates by angle equal to the doubled phase. The 2D flickering (and respective 3D rotation) produces no damage to the involved algebras.*

The simplest transformation of the type (9) for the dyad (8) is the "rotation" about vector $\mathbf{q}_{\tilde{3}}$ at angle $\alpha$, the rotation matrix is $S = \cos\alpha + \mathbf{q}_3 \sin\alpha$; the results of the transformation are

$$\psi^\pm = (\cos\alpha \pm i \sin\alpha)\tilde{\psi}^\pm = e^{\pm i\alpha}\tilde{\psi}^\pm , \quad \varphi^\pm = e^{\mp i\alpha}\tilde{\varphi}^\pm . \tag{10}$$

Eqs.(10) state that real and imaginary constituents of the dyad vectors $\psi^\pm$ length harmonically change with $\alpha$ so that an area of the fractal space formed by the vectors is "pumped over" (flickers) from real sector to imaginary sector of the 2D-cell (the same with covectors $\varphi^\pm$). Using Eqs.(6) we compute the results of respective transformations of the algebraic units

$$1 = \tilde{1}, \quad \mathbf{q}_1 = \mathbf{q}_{\tilde{1}} \cos 2\alpha + \mathbf{q}_{\tilde{2}} \sin 2\alpha , \quad \mathbf{q}_2 = \mathbf{q}_{\tilde{1}} \cos 2\alpha - \mathbf{q}_{\tilde{2}} \sin 2\alpha , \quad \mathbf{q}_3 = \mathbf{q}_{\tilde{3}}, \tag{11}$$

i.e. the scalar unit does not change, while the 3D frame is rotated about vector $\mathbf{q}_3 = \mathbf{q}_{\tilde{3}}$ at $2\alpha$ (twice as spinor vectors "rotation"); the units (11) well fit for all involved algebras, this may be verified by direct computation.

(vi) *The flickering 2D-cell can be stretched ("loaded with a fractal density"); this transformation causes a 2D metric defect, changes lengths of the rotated 3D frame vectors thus damaging the algebras.*

Let the flickering dyad (10) be subject to an extra transformation, conformal stretching

$$\psi'^{\pm} = \sigma e^{\pm i\alpha}\tilde{\psi}^{\pm} \equiv \lambda \tilde{\psi}^{\pm}, \quad \varphi'^{\pm} = \sigma e^{\mp i\alpha}\tilde{\varphi}^{\pm} \equiv \lambda^* \tilde{\varphi}^{\pm}, \qquad (12)$$

$\sigma \in \mathbf{R}$. The mapping (12) injects the 2D-cell's metric defect

$$1' = \psi'^{+}\varphi'^{+} + \psi'^{-}\varphi'^{-} = \sigma^2, \qquad (13)$$

all vector units acquiring the same square factor, e.g. $\mathbf{q}_{3'} = \sigma^2 \mathbf{q}_3$; this evidently damages the multiplication law (2), so all involved algebras are violated. In 3D space $\sigma^2$ may be thought of as a density, so the factor $\sigma$ will be called "fractal density" loading formerly "empty" 2D-cell.

### 3.3 An abstract (exterior) space and the algebras' stability condition

(vii) *A normalizing integral (a functional) of a dyad vector square length over a volume of an abstract M-dimensional space (in particular, an abstract 3D space) smoothes the 2D metric defect down, retuenes the unit lengths of 3D frame vectors thus restoring the algebras.*

The metric defect (13) is smoothed down if the factor

$$\lambda = \sigma e^{i\alpha} \qquad (14)$$

is a compact function $\lambda(\xi_\Lambda, \theta)$ (of coordinates $\xi_\Lambda$, $\Lambda = 1, 2 \ldots M$ of $M$-dimensional abstract space and a free parameter $\theta$, all magnitudes measured in no units) in a volume $V_\Lambda$

$$f \equiv \int_{V_\Lambda} \varphi'^{\pm}\psi'^{\pm}dV_\Lambda = \int_{V_\Lambda} \lambda\lambda^* dV_\Lambda = \int_{V_\Lambda} \sigma^2 dV_\Lambda = 1. \qquad (15)$$

Then the objects built out of dyad (12) as in Eq.(6), but "seen" from the space, do not differ from those of Eq. (11) and can serve as good algebra units, e.g.

$$1' = f(\psi'^{+}\varphi'^{+} + \psi'^{-}\varphi'^{-}) = \tilde{1}, \quad \mathbf{q}_{3'} = if(\psi'^{+}\varphi'^{+} - \psi'^{-}\varphi'^{-}) = \mathbf{q}_{\tilde{3}}, \qquad (16)$$

$$\mathbf{q}_{1'} = -if(\psi'^{+}\varphi'^{-} + \psi'^{-}\varphi'^{+}) = (\cos 2\alpha)\mathbf{q}_{\tilde{1}} + (\sin 2\alpha)\mathbf{q}_{\tilde{2}} = \mathbf{q}_1.$$

(viii) *The algebras are saved "forever" in the sense of the free parameter $\theta$ if the normalizing functional is constant with respect to the parameter's change; this condition of the algebras' stability entails a continuity-type equation for square of the fractal density.*

The normalization (15) "lasts forever" in the sense of $\theta$ (thus providing the algebras' stability) if the function $\lambda\lambda^*$ satisfies the continuity-type equation

$$\partial_\theta(\lambda\lambda^*) + \partial_\Lambda(\lambda\lambda^* k_\Lambda) = 0, \qquad (17)$$

where $\partial_\theta \equiv \partial/\partial\theta$, $\partial_\Lambda \equiv \partial/\partial\xi_\Lambda$, summation in index $\Lambda$ is implied, $k_\Lambda$ is a vector of the 2D-cell "propagation" in the abstract space.

### 3.4. Basic fractal equations, consequences of the stability condition

(ix) <u>Schrodinger-type equation</u>. *If propagation vector of the 2D-cell (in the abstract space) is just a gradient of the flickering phase, then the continuity-type equation decays into a couple of mutually conjugate equations each mathematically equivalent to the Schrodinger equation of quantum mechanics.*

Let propagation of 2D-cell be determined by the phase increase. The phase is expressed from Eq.(14) $\alpha = \frac{i}{2}\ln\frac{\lambda^*}{\lambda}$, then the propagation vector $k_\Lambda = \partial_\Lambda \alpha = \frac{i}{2}\left(\frac{\partial_\Lambda \lambda^*}{\lambda^*} - \frac{\partial_\Lambda \lambda}{\lambda}\right)$ when inserted in Eq.(17) brings it to the form

$$\left[\partial_\theta - \frac{i}{2}(\partial_\Lambda \partial_\Lambda - 2W)\right]\lambda + e^{2\alpha}\left[\partial_\theta + \frac{i}{2}(\partial_\Lambda \partial_\Lambda - 2W)\right]\lambda^* = 0,$$

where $W(\xi,\theta)$ is an arbitrary real function. If the last equation holds for all values of the phase then each of its conjugate parts should vanish. Equation for $\lambda$

$$\left[\partial_\theta - \frac{i}{2}(\partial_\Lambda \partial_\Lambda - 2W)\right]\lambda = 0 \qquad (18)$$

is an exact math analogue of Schrodinger equation of quantum mechanics.

(x) _Pauli-type equation_. *If propagation vector of the 2D-cell apart from the phase's gradient includes an exterior vector field, then the continuity-type equation decays into mutually conjugate math equivalents of the Pauli equation of quantum mechanics.*

Consider and a more general case of the 2D-cell propagation vector $k_n = \partial_n \alpha + A_n$ where $A_k(x_n, t)$ is some exterior vector field (for simplicity the abstract space here is chosen three-dimensional with coordinates $\xi_n$). Presence of the vector field induces return to full spinor functions in the normalizing integral $\int_{V_n} \varphi' \psi' dV_n = 1$, (here $\varphi', \psi'$ and other spinors are chosen e.g. of positive parity), and representation of the 3D space metric in the Clifford algebra format $\delta_{kn} \equiv \frac{1}{2}(\mathbf{p}_k \mathbf{p}_n + \mathbf{p}_n \mathbf{p}_k)$, where $\mathbf{p}_k \equiv i\mathbf{q}_k$. Respective continuity-type equation is written as $\partial_\theta(\widetilde{\varphi}\lambda^*\lambda\widetilde{\psi}) + \frac{1}{2}(\mathbf{p}_m \mathbf{p}_n + \mathbf{p}_n \mathbf{p}_m)\partial_m[\widetilde{\varphi}\lambda^*\lambda\widetilde{\psi}(\partial_n \alpha + A_n)] = 0$, it decays into mutually Hermitian conjugate parts; equation for function $\psi'$ has the form

$$\left[i\partial_\theta - \frac{1}{2}(-i\partial_k + A_k)(-i\partial_k + A_k) - \frac{1}{2}\mathbf{p}_k B_k - W\right]\psi' = 0, \qquad (19)$$

where $B_k \equiv \varepsilon_{kmn}\partial_m A_n$. Eq.(25) is a math analogue of Pauli equation of quantum mechanics describing motion of electron in exterior magnetic field. Details of computation of Eq.(19) are found in ref. [9].

(xi) _Klein-Gordon-type equation_. *If the exterior space is a Minkowsky-type space-time with $\theta$ being time-like coordinate, and propagation vector of the 2D-cell being a 4D gradient of the phase, then the continuity-type equation decays into mutually conjugate math equivalents of the Klein-Gordon-type equation.*

Let the exterior space be formally represented as a "space-time" having the indefinite metric $\delta^{\mu\nu} = diag\,(1,-1,-1,...-1)$ and coordinates $\xi^\mu$, $\mu,\nu... = 0,1,2,...M$ with $\xi^0 \equiv \theta$; let also the complex factor $\lambda = \sigma e^{i\alpha}$ be "re-gauged" (shown as a product of two complex numbers) $\lambda(\xi^\mu) = \gamma\overline{\lambda}$ where $\gamma\gamma^* \equiv \overline{k}_0$, and $\overline{\lambda} \equiv \overline{\sigma}e^{i\Phi}$. Then, defining $(M+1)$-dimensional propagation vector $\overline{k}^\mu = \{\overline{k}^0, \overline{k}^\Lambda\}$, where $\overline{k}^\Lambda \equiv \overline{k}_0 k_\Lambda$, $k_\Lambda$ being the propagation vector from Eq.(17), we can identically rewrite this continuity-type equation in the form

$$\partial_\mu(\overline{\lambda}\overline{\lambda}^*\,\overline{k}^\mu) = 0. \qquad (20)$$

Next, we demand that the propagation vector be gradient of the phase $\overline{k}^\mu = \delta^{\mu\nu}\partial_\nu \Phi = \frac{i}{2}\delta^{\mu\nu}\partial_\nu \ln\frac{\overline{\lambda}^*}{\overline{\lambda}}$, then Eq.(20) quadratic in $\overline{\lambda}$ decays into a conjugate couple of the Klein-Gordon-type equations, linear in the factor, that for $\overline{\lambda}$ is

$$(\delta^{\mu\nu}\partial_\mu\partial_\nu - \overline{W})\overline{\lambda} = 0, \tag{21}$$

here $\overline{W}(\xi^\nu)$ is an arbitrary function. Under simple conditions Eq.(21) is reduced to the Schrodinger-type equation (18). Indeed, put the function $\overline{\lambda}$ in the form

$$\overline{\lambda} = \gamma^{-1}\lambda \equiv \varsigma e^{i\eta}\lambda, \tag{22}$$

with $\varsigma = 1 + o_1$, $\eta = \theta + o_2$, where $o_1$, $o_2$ are small functions as well as all their derivatives. Insert the function (22) into Eq.(21); straightforward computations yield the sought for result

$$\left[i\partial_0 - \frac{1}{2}\partial_\Lambda\partial_\Lambda + \frac{1}{2}(-\overline{W}-1)\right]\lambda = 0$$

i.e. Eq.(18) with $\overline{W} + 1 \equiv -2W$.

### 3.5. Transition to the physical space and introduction of the scales

(xii) *All above math equations contain magnitudes measured in no physical units; transition from an abstract to the physical space compels to introduce space-time standards. Short scale standards are chosen.*

Instead of $M$-dimensional abstract space (where $\xi_\Lambda$ and $\theta$ are measured in no units) the math equations can be regarded over 3D physical space and time necessarily scaled $\xi_\Lambda \to x_k/\varepsilon$, $\theta \to t/\tau$, where $\varepsilon$ and $\tau$ are space-length and time-interval standards. Characteristic space length is chosen equal to the Compton wave length

$$\varepsilon \equiv \frac{\hbar}{mc}, \tag{23a}$$

where $\hbar$ is the Planck constant, $c$ is velocity of light in vacuum, $m$ is the electron's rest mass (in these constants the space scale is assessed as $\varepsilon \cong 10^{-11} cm$). Respective time standard is the time interval needed for light to travel (in vacuum) along the characteristic length

$$\tau \equiv \frac{\varepsilon}{c} = \frac{\hbar}{mc^2} \tag{23b}$$

(the time scale is assessed as $\tau \cong 10^{-22} s$).

(xiii) *In these units the 2D-cell describes a pre-geometric protoparticle its fractal density function acquiring the sense of relative fractal mass density (the density function per mean density), so that the normalizing functional (15) converts into definition of mass of a 3D particle (based on a 2D protoparticle).*

The function $\sigma$ remains unit-less (measured in no units), so it may have sense of a "relative fractal mass density" $\sigma \equiv \sqrt{\rho(x,t)/\rho_{mean}}$, where $\rho_{mean}$ is the mean mass density of the particle (electron) in the 3D volume it is supposed to occupy. Now a model of "protoparticle" emerges. As a fractal object it is conceived as a $\sigma$-weighted 2D-cell $\{\sigma e^{i\alpha}\widetilde{\psi}^\pm\}$; with change of the phase its area (hence, weight) is flickering between real and imaginary sectors. In fact it is a "visual image" of the particle's state (wave) function of quantum mechanics. The normalizing integral (15) is converted to definition of the particle's mass

$$\frac{1}{\varepsilon^3}\int_V \sigma^2(x,t)dV = 1 \quad \to \quad \int_V \rho(x,t)dV = \varepsilon^3 \rho_{mean} = m. \tag{24}$$

### 3.6. In the physical space the math equations become physical laws

(xiv) *In the chosen physical units Eq.(18) and Eq.(19) become respectively exact Schrodinger and Pauli equations, Eq.(21) becomes extended Klien-Gordon equation.*

One can easily verify that in coordinates and time scaled as in Eqs.23, Eq.18 takes the form of Schrodinger equation

$$\left(i\hbar\partial_t + \frac{\hbar^2}{2m}\partial_k\partial_k - U\right)\lambda(x,t) = 0, \qquad (25)$$

where $U \equiv mc^2 W$ is a scalar potential; Eq.(19) takes the form of Pauli equation

$$\left[i\hbar\partial_t - \frac{1}{2m}(-i\hbar\partial_k + \frac{q}{c}\widetilde{A}_k)(-i\hbar\partial_k + \frac{q}{c}\widetilde{A}_k) - \frac{q\hbar}{2mc}\mathbf{p}_k\widetilde{B}_k - U\right]\Psi(x,t) = 0, \qquad (26)$$

where $q$ is the electric charge, $\widetilde{A}_k \equiv \frac{mc^2}{q}A_k$, $\widetilde{B}_k \equiv \frac{mc^2}{q}B_k$ are potential and intensity of the magnetic field respectively and $U \equiv mc^2 W$ is a scalar potential. Eq.(21) in the chosen physical units becomes the extended Klien-Gordon equation

$$\left[\hbar^2\left(\frac{1}{c^2}\partial_t\partial_t - \partial_n\partial_n\right) - m^2c^2(1+W')\right]\lambda(x,t) = 0, \qquad (27)$$

the free function represented in the form $1+W' \equiv \overline{W}$; Eq.27 obviously admits fractalization, e.g. the Dirac square root format.

### 3.7. Hamilton-Jacobi equation.

(xv) *Any Schrodinger-type math equation has complex-number structure; it can be separated into real and imaginary parts; the real part is a math equivalent of conservation law for semi-density function.*

Return to the Schrodinger-type equation (18) and taking into account $\lambda = \sigma e^{i\alpha}$ decompose it into real and imaginary parts thus obtaining the system of Bohm-type equations [10]

$$\partial_\theta\sigma + \partial_\Lambda\sigma\partial_\Lambda\alpha + \frac{1}{2}\sigma\partial_\Lambda\partial_\Lambda\alpha = 0, \qquad (28)$$

$$\partial_\theta\alpha + \frac{1}{2}(\partial_\Lambda\alpha)(\partial_\Lambda\alpha) + W - \frac{1}{2}\partial_\Lambda\partial_\Lambda\sigma/\sigma = 0. \qquad (29)$$

The real component (28) multiplied by $\sigma$ is converted into density conservation-type equation

$$\partial_\theta\sigma^2 + \partial_\Lambda(\sigma^2\partial_\Lambda\alpha) = 0. \qquad (30)$$

(xvi) *The imaginary part is a math equivalent of the Hamilton-Jacobi equation of classical mechanics, the 2D-cell's flickering phase (or angle of the 3D frame rotation) playing the role of the action function.*

The imaginary component (29) of the Bohm-type system has the form

$$\partial_\theta\alpha + \frac{1}{2}(\partial_\Lambda\alpha)(\partial_\Lambda\alpha) + W - \frac{1}{2}\partial_\Lambda\partial_\Lambda\sigma/\sigma = 0. \qquad (31)$$

If all terms in Eq.(31) are fast changing functions ("inside" the 2D-cell), then the system of Eqs.(30, 31) is just an equivalent of Eq.(18). But it may happen that only the last term in Eq.(31) (depending on $\sigma$) is fast changing one (short-scale), while the terms depending on $\alpha$ are functions slowly changing "outside" the 2D-cell (long-scale). In this case we have to consider the free function $W = W_{in} + W_{ex}$ as split into "interior" and "exterior" parts; so Eq.(30) decays into respective equations

$$\partial_\Lambda \partial_\Lambda \sigma - 2W_{in}\sigma = 0. \tag{32}$$

$$\partial_\theta \alpha + \frac{1}{2}(\partial_\Lambda \alpha)(\partial_\Lambda \alpha) + W_{ex} = 0. \tag{33}$$

Static Eq.(32) determines the fractal-density distribution under influence of some interior reason $W_{in}$. But Eq.(33) is the familiar math analogue of Hamilton-Jacobi equation of classical mechanics, the phase $\alpha$ of 2D-cell flickering (or half angle of the frame $\mathbf{q}_k$ rotation about $\mathbf{q}_3$) playing role of the action function, the term $W_{ex}$ behaving as an exterior potential.

(xvii) *In physical units (on laboratory scale) the phase of 2D-cell's flickering is a mechanical action function measured in of the Planck constant. Then the dynamic math equations following Bohm-type equations become the mass conservation equation and Hamilton-Jacobi equation. Static equation determins the fractal mass density distribution.*

Let the phase be a slowly changing (laboratory scale) function in the chosen above physical units. Then $\partial_\Lambda \to (\hbar/mc)\partial_n$, $\partial_\theta \to (\hbar/mc^2)\partial_t$ and taking into account that $\sigma^2 \sim \rho$ Eq.(30) is rewritten as

$$\partial_t \rho + \partial_n(\rho u_n) = 0 \tag{34}$$

where $u_k \equiv \partial_k S/m$ is 3D velocity, $S$ being the classical mechanical action function is the phase measured in the Planck constant

$$S(x,t) \equiv \hbar \alpha(x,t). \tag{35}$$

Eq.(31) in the physical units becomes precisely the Hamilton-Jacobi equation

$$\partial_t S + \frac{1}{2m}(\partial_m S)(\partial_m S) + U_{ex} = 0, \tag{36}$$

where $U_{ex} \equiv mc^2 W_{ex}$. We emphasize that Eq.(36) deduced in fact as a square root from 3D continuity-type equation should be referred to as a spinor (fractal) equation of classical mechanics.

Finely, Eq.(32) in the physical units has the form

$$\partial_m \partial_m \sigma - R_{in}\sigma = 0, \tag{37}$$

where $R_{in} \equiv 2W_{in}/\varepsilon^2$ is some interior "potential" measured (as e.g. curvature) in $cm^{-2}$; we leave it for future explorations.

### 3.8. Geometric (physical) equations

(xviii) *The flickering phase's minimal value on the free parameter segment entails a math equivalent of the Euler-Lagrange equation (Newton's dynamic law) of classical mechanics.*

Back to math equations (33), let us replace in it the partial derivative by the full one $\partial_\theta \alpha = d_\theta \alpha - d_\theta \xi_\Lambda \cdot \partial_\Lambda \alpha$ to obtain the phase value integral on the segment $[\theta_1, \theta_2]$

$$\alpha = \int_{\theta_1}^{\theta_2} \left( d_\theta \xi_\Lambda \, \partial_\Lambda \alpha - \frac{1}{2}\partial_\Lambda \alpha \, \partial_\Lambda \alpha - W_{ext} \right) d\theta. \tag{34}$$

The "minimal phase" demand selects extreme lines $\xi_\Lambda(\theta)$ with "observables" $d_\theta \xi_\Lambda$ (M-dimensional velocity) and $\partial_\Lambda \alpha$ (momentum) obeying the equation

$$\partial_\theta \left[ \partial_K \alpha + \frac{\partial(\partial_\Lambda \alpha)}{\partial(d_\theta \xi_K)} (d_\theta \xi_\Lambda - \partial_\Lambda \alpha) \right] + \partial_K W_{ext} = 0. \tag{35}$$

We recognize in the integrand in Eq.(34) a math analogue of Lagrangian function of classical mechanics, Eq.(35) is a math analogue of dynamic equation of Newtonian dynamics. In the physical units the derivatives are $d_\theta \xi_\Lambda \to \tau d_t(x_n/\varepsilon) = u_n/c$, $\partial_\Lambda \alpha \to \varepsilon \partial_n(S/\hbar) = u_n/c$; their insertion in Eq.(34, 35) converts them respectively into the classical action functional

$$S = \int_{t_1}^{t_2} \left( \frac{1}{2} m u_n u_n - U_{ext} \right) dt \tag{36}$$

where $U_{ext} \equiv mc^2 W_{ex}$, and the Newtonian dynamic equation

$$\partial_t(m u_n) + \partial_n U_{ext} = 0. \tag{37}$$

Strange enough, this basic physical law (37) discovered in experiment appears is just a very special case of a simple purely mathematical model.

3.9. The helix model and relativistic particle

(xix) *The model of a particle in the 3D space is a point-like mass (distributed in a very small volume of the characteristic length size) with a frozen-in 3D frame able to rotate. If it is permanently rotating then at a point of the particle's ultimate radius (half of the scale unit) velocity of rotation equals that of light; if this particle moves then this point depicts in the space a cylindrical helix line. Velocity of moving particle's border point remains maximal, i.e. that of light.*

Geometrically the particle is conceived as a mass $m$ distributed in small 3D volume of the size $\varepsilon$ with a triad $\mathbf{q}_k$ "frozen" in its center and rotated (with the mass) about one its vector by angle equal to doubled 2D-cell's flickering phase; the angle's gradient points direction of the particle's motion. A free particle moving in 3D space along coordinate $z$ with velocity $dz/dt \equiv u = const$, ($t$ is the observer's time) satisfies two conditions:

  (1) the particle's triad rotates permanently about $\mathbf{q}_3$ with frequency $d(2\alpha)/dt \equiv 2\omega = const$, the angle between $\mathbf{q}_3$ and the velocity vector is $\beta = const$;
  (2) a point at the particle's ultimate radius $\varepsilon/2$ depicts a helix-type line, the point's linear velocity is always maximal, i.e. that of light.

(xx) *The difference between squares of the free particle's helix small length and path has the form of the space-time interval of the Special Relativity; computed in physical units this interval gives the action function of a relativistic particle.*

The helix line given by the coordinate functions $x = (\varepsilon/2)\cos 2\alpha \cos\beta$, $y = (\varepsilon/2)\sin 2\alpha$, $z = ut - (\varepsilon/2)\cos 2\alpha \sin\beta$, has the line element $dl^2 = \varepsilon^2 d\alpha^2 + 2\varepsilon \sin 2\alpha \sin\beta \, d\alpha \, udt + u^2 dt^2$. Condition (2) means $dl = cdt$, then from the line-element we obtain

$$c^2 = \varepsilon^2 \omega^2 + 2u\varepsilon\omega \sin(2\omega t)\sin\beta + u^2. \tag{38}$$

For moving free particle ($u = const$) Eq.(38) holds if $\beta = 0$, i.e. the regular helix lies on the circular cylinder $dl^2 = c^2 dt^2 = \varepsilon^2 d\alpha^2 + u^2 dt^2$; find the 2D-cell's phase on the segment $[t_1, t_2]$

$$\alpha = \pm \frac{c}{\varepsilon} \int_{t_1}^{t_2} \sqrt{1 - \frac{u^2}{c^2}} \, dt \, , \tag{39}$$

the signs indicating right or left helicity. Insert $\varepsilon = \hbar/(mc)$ and chose the sign "minus", then, provided Eq.(32) is taken into account, Eq.(39) yields action of a free relativistic particle

$$S = \alpha \hbar = -mc^2 \int_{t_1}^{t_2} \sqrt{1 - \frac{u^2}{c^2}} \, dt \, . \tag{40}$$

So the line element of special relativity $\varepsilon^2 d\alpha^2 \equiv ds^2 = c^2 dt^2 - dz^2$ acquires a specific geometric meaning. The "space-time interval" $ds = \varepsilon d\alpha$ has the sense of an arc length of the particle's "ultimate circumference", the immobile particle's border in the plane of rotation. For a free particle in its own frame this arc length is a definite unchanging number (invariant of special relativity).

(xxi) *Reduction to the non-relativistic case automatically establishes relations between classical and quantum magnitudes, thus determining the free particle's 2D model (protoparticle) as De Broglie wave, the particle's rest energy linked with 2D-cell's permanent flickering.*

We rewrite Eq.(40) in differential form and reduce it to the non-relativistic case

$$\hbar d\alpha \cong -E \, dt + p_n dx_n \equiv -(mc^2 + mu^2/2) dt + mu_n dx_n \cong dS \, , \tag{41}$$

the free particle is considered to be a quantum one, $\hbar d\alpha = \hbar \partial_t \alpha \, dt + \hbar \partial_n \alpha \, dx_n$. Then from Eq.(41) we have $\hbar \omega dt + \hbar k_n dx_n \cong -E dt + p_n dx_n$ thus automatically obtaining De Broglie's energy-frequency and momentum-wave vector ratios

$$E = (mc^2 + mu^2/2) = |\omega|\hbar \, , \quad p_n = k_n \hbar \, ; \tag{40}$$

the state function of the particle (e.g. with positive parity)

$$\psi'^+ = \sigma e^{i\alpha} \widetilde{\psi}^+ = \sigma e^{i(p_n x_n - Et)/\hbar} \widetilde{\psi}^+ = \sigma e^{i(k_n x_n - \omega t)} \widetilde{\psi}^+ \tag{41}$$

has precise form of the De Broglie wave. From Eqs.(40) one in particular notes that even immobile in space ($u = 0$) the particle's 2D-cell must be permanently "pumped over" ($\omega \neq 0$), the flickering frequency $|\dot{\alpha}| \equiv \omega_0$ determining the particle's rest energy $E_{rest} \equiv mc^2 = \omega_0 \hbar$.

(xxii) *In the classical case the helix model expectedly yields Hamilton-Jacobi equation.*

Let the free particle be classical with $dS = \partial_t S + u_n \partial_n S$. Then from Eq.(39) we have $-mc^2 + mu^2/2 = \partial_t S + u_n \partial_n S$, or Hamilton-Jacobi equation $\partial_t S + \partial_n S \partial_n S /(2m) + E_{rest} = 0$, here appearing as a consequence of the helix model (no exterior potentials for a free particle).

### 3.10. Irregular helix model and "more general" relativity

(xxiii) *Making the 3D particle's helix line irregularly curved and compressed (as if the motion proceeds under influence of a variable exterior force) leads to appearance of variable metric components and general relativity-type "space-time" line element, the space components though being negligibly small on laboratory scale. A metric function measuring the helix compression emerges in respective Hamilton-Jacobi equation as an exterior potential.*

A point of particle's triad vector (e.g. of $\mathbf{q}_1$) in a force field must depict a distorted helix line, its length element determined as follows. Let $\mathbf{q}_n(t,x)$ be the particle's Frenet-type triad with $\mathbf{q}_3$ tangent to the particle's trajectory $x_k(t)$, its local curvature being $R(x)$. Define a quaternion radius vector $\mathbf{l} \equiv \varepsilon \mathbf{q}_1 + r\mathbf{q}_3$ where $\varepsilon$ is the helix diameter (constant), $r$ is a length along the trajectory; the differential of $\mathbf{l}$ is

$$d\mathbf{l} \equiv r\omega_{331} dr\, \mathbf{q}_1 + (\varepsilon\omega_{312} + r\omega_{332}) dr\, \mathbf{q}_2 + (1 - \varepsilon\omega_{331}) dr\, \mathbf{q}_3,$$

since $d\mathbf{q}_n = \omega_{jnk} dx_j \mathbf{q}_k$ where $\omega_{nkj}$ are connection components, among them $\omega_{312} = d\alpha/dr$ is "torsion" (rotation of $\mathbf{q}_1$ about $\mathbf{q}_3$), $\omega_{331} \equiv R$ is the trajectory's first curvature, and $\omega_{332}$ (neglected for simplicity) is the second curvature [5]. So the "curved" helix line element has the form

$$dl^2 = d\mathbf{l}\, d\bar{\mathbf{l}} = \varepsilon^2 d\alpha^2 + [(1-\varepsilon R)^2 + z^2 R^2] dr^2 \equiv \varepsilon^2 d\alpha^2 + e^{2G(x)} dr^2.$$

If the helix is additionally "compressed" with the measure $e^{-2W(x)}$ then the line element becomes

$dl^2 = e^{-2W(x)}[\varepsilon^2 d\alpha^2 + e^{2G(x)} dr^2]$, and the helix "space-time interval" acquires the features of general relativity

$$\varepsilon^2 d\alpha^2 = ds^2 = e^{2W(x)} c^2 dt^2 - e^{2G(x)} dr^2. \tag{42}$$

For small spatial curvatures $e^{2G} \approx 1$, and for non-relativistic classical particle Eq.(42) gives the action differential ($dS = \hbar d\alpha$) as $dS = -mc^2 dt \sqrt{e^{2W} - (u/c)^2} \approx -[mc^2(1+W) - mu^2/2] dt$, or equivalently, the Hamilton-Jacobi equation $\partial_t S + \partial_n S \partial_n S /(2m) + U = 0$ with the potential $U \equiv mc^2 W + E_{rest}$.

(xxiv) *The 3D space Euler-Lagrange equation of a "squeezed helix particle" exactly coincides with the 4D space-time geodesic equation, thus demonstrating general relativity and the helix model theory convergence.*

Variation of the "space-time" interval $\delta \int_a^b \sqrt{g_{\mu\nu} u^\mu u^\nu}\, ds = 0$, $u^\mu \equiv dx^\mu/ds$ of the compressed helix $ds^2 = g_{00}(dx^0)^2 - \delta_{kn} dx^k dx^n = e^{2W(x)} c^2 dt^2 - dr^2$ yields equation of extremal (geodesic) line $d_s(g_{\mu\lambda} u^\lambda) = g_{\alpha\beta} u^\alpha u^\beta /2 \rightarrow m\partial_t u_k = -\partial_k U(e^{2W} \delta_{kn} - 2u_k u_n /c^2)$, the same as the Euler-Lagrange equation following from the Lagrangian $L(x,\dot{x}) = -mc^2 \sqrt{e^{2W(x)} - \dot{x}_k \dot{x}_k /c^2}$. So the "irregular relativistic helix" model, a consequence of the 2D-cell conjecture, partially explains heuristic geometrization of interactions.

## 4. Conclusion

Here we summarize the logic and structure of the suggested theory. There are two parallel but drastically different realms in the theory, one is an "unobservable" area comprising primitive math relations and pregeometric images on the fractal surface, the other is an "observable" area containing equalities composed of the primitive ones, and usual geometric objects in physical space. We start with deformations of a small pregeometric domain in the same time trying not to wreck properties of geometric objects thus saving a set of algebras; the price is a series of fractal equations. Written in physical units these pure math equalities lead precisely to equations of quantum mechanics (Schrodinger, Pauli) and of classical mechanics (Hamilton-Jacobi equation, Newtonian dynamic equation). Simultaneously a fractal protopartical model arises, its phase having sense of mechanical action; respective geometric analog is a rotating massive point-like particle. This model leads to an original "helix-line" formulation of mechanics of a free relativistic particle (if the helix is a regular cylindrical "spring"; for an irregular "spring" the space-time metric becomes point dependent, and the relativistic mechanics is described by geodesic equation, so that the theory acquire features of general relativity. Respective non-relativistic fractal equation is again that of Hamilton-Jacobi.

There are some challenging problems waiting for solution; among them analysis of the static equation a particle's fractal density (as well as physical density) distribution, construction of pregeometric models for massless particles, and possibly for electrical charge.

**References**
1. A.P.Yefremov, Structured relativistic particle, helix-type Minkowski diagram, and more general relativity, Gravit. and Cosmol. **20** (3),226 (2014).
2. A.P.Yefremov, Structure of hypercomplex units and exotic numbers as sections of bi-quaternions, Adv.Sci.Lett. **3**, 537 (2010).
3. V.V.Kisil, Erlangen program at large – 2: inventing a wheel. The parabolic one, math.GM/07074024v1.
4. P.Fjelstad, Extending special relativity via the perplex numbers. Am.J.Phys. **54 (**5), 416 (1986)**.**
5. A.P.Yefremov, Quaternion Spaces, Frames and Fields, (PFUR publ., Moscow, 2005).
6. A.P.Yefremov, The conic-gearing image of a complex number and a spinor-born surface geometry, Gravit. and Cosmol. **17** (1), 1 (2011).
7. W.R.Hamilton, The Mathematical Papers of William Rouan Hamilton. App. 3, Vol. 3 (Cambrige: CUP, 1967).
8. J. A. Wheeler, Pregeometry: motivations and prospects. In: A. R. Marlov (ed.), Quantum Theory and Gravitation (New York, Academic Press, 1080) pp. 1–11.
9. A.P.Yefremov, Pre-geometric structure of quantum and classical particles in terms of quaternion spinors, Gravit. and Cosmol. **19** (2), 71 (2013).
10. D.Bohm, A Suggested Interpretation of the Quantum Theory in Terms of "Hidden" Variables. I. Phys.Rev. **85**, 166 (1952**)**.